\documentclass[journal]{IEEEtran}
\usepackage{amsmath,amsfonts}
\usepackage{algorithmic}
\usepackage{algorithm}
\usepackage{array}
\usepackage[caption=false,font=normalsize,labelfont=sf,textfont=sf]{subfig}
\usepackage{textcomp}
\usepackage{stfloats}
\usepackage{url}
\usepackage{verbatim}
\usepackage{graphicx}
\usepackage{cite}
\usepackage{gensymb}
\newcommand{\lt}{\ensuremath{<}}

\usepackage{placeins}
\usepackage{afterpage}

\usepackage{aas_macros}

\usepackage[table]{xcolor}
\usepackage{nicematrix}
\definecolor{mygray}{gray}{0.90}
\definecolor{myblue}{RGB}{220,235,245}
\definecolor{headgray}{gray}{0.92}
\definecolor{mywhite}{rgb}{1,1,1}

\begin{document}

\title{Design and Performance of the SPT-SLIM Receiver Cryostat}

\author{%
\IEEEauthorblockN{
M.~R.~Young\IEEEauthorrefmark{1}\IEEEauthorrefmark{2},
M.~Adamic\IEEEauthorrefmark{3},
A.~J.~Anderson\IEEEauthorrefmark{1}\IEEEauthorrefmark{2}\IEEEauthorrefmark{4},
P.~S.~Barry\IEEEauthorrefmark{5},
B.~A.~Benson\IEEEauthorrefmark{1}\IEEEauthorrefmark{2}\IEEEauthorrefmark{4},
C.~S.~Benson\IEEEauthorrefmark{5},
E.~Brooks\IEEEauthorrefmark{4},
J.~E.~Carlstrom\IEEEauthorrefmark{2}\IEEEauthorrefmark{6}\IEEEauthorrefmark{7}\IEEEauthorrefmark{8}\IEEEauthorrefmark{4},
T.~Cecil\IEEEauthorrefmark{8},
C.~L.~Chang\IEEEauthorrefmark{8}\IEEEauthorrefmark{2}\IEEEauthorrefmark{4},
K.~R.~Dibert\IEEEauthorrefmark{4},
M.~Dobbs\IEEEauthorrefmark{3},
K.~Fichman\IEEEauthorrefmark{7}\IEEEauthorrefmark{2},
M.~Hollister\IEEEauthorrefmark{1},
K.~S.~Karkare\IEEEauthorrefmark{9},
G.~K.~Keating\IEEEauthorrefmark{10},
A.~M.~Lapuente\IEEEauthorrefmark{9},
M.~Lisovenko\IEEEauthorrefmark{8},
D.~P.~Marrone\IEEEauthorrefmark{11},
D.~Mitchell\IEEEauthorrefmark{1},
J.~Montgomery\IEEEauthorrefmark{3},
T.~Natoli\IEEEauthorrefmark{2},
Z.~Pan\IEEEauthorrefmark{8}\IEEEauthorrefmark{2}\IEEEauthorrefmark{7},
A.~Rahlin\IEEEauthorrefmark{4}\IEEEauthorrefmark{2},
G.~Robson\IEEEauthorrefmark{5},
M.~Rouble\IEEEauthorrefmark{3},
G.~Smecher\IEEEauthorrefmark{12}\IEEEauthorrefmark{3},
V.~Yefremenko\IEEEauthorrefmark{8},
C.~Yu\IEEEauthorrefmark{8}\IEEEauthorrefmark{2}\IEEEauthorrefmark{4},
J.~A.~Zebrowski\IEEEauthorrefmark{2}\IEEEauthorrefmark{4}\IEEEauthorrefmark{1},
C.~Zhang\IEEEauthorrefmark{13},
}\\
\IEEEauthorblockA{\IEEEauthorrefmark{1}Fermi National Accelerator Laboratory, MS209, P.O. Box 500, Batavia, IL, 60510, USA}\\
\IEEEauthorblockA{\IEEEauthorrefmark{2}Kavli Institute for Cosmological Physics, University of Chicago, 5640 South Ellis Avenue, Chicago, IL, 60637, USA}\\
\IEEEauthorblockA{\IEEEauthorrefmark{3}Department of Physics and Trottier Space Institute, McGill University, 3600 Rue University, Montreal, Quebec H3A 2T8, Canada}\\
\IEEEauthorblockA{\IEEEauthorrefmark{4}Department of Astronomy and Astrophysics, University of Chicago, 5640 South Ellis Avenue, Chicago, IL, 60637, USA}\\
\IEEEauthorblockA{\IEEEauthorrefmark{5}School of Physics and Astronomy, Cardiff University, Cardiff CF24 3YB, United Kingdom}\\
\IEEEauthorblockA{\IEEEauthorrefmark{6}Enrico Fermi Institute, University of Chicago, 5640 South Ellis Avenue, Chicago, IL, 60637, USA}\\
\IEEEauthorblockA{\IEEEauthorrefmark{7}Department of Physics, University of Chicago, 5640 South Ellis Avenue, Chicago, IL, 60637, USA}\\
\IEEEauthorblockA{\IEEEauthorrefmark{8}High-Energy Physics Division, Argonne National Laboratory, 9700 South Cass Avenue., Lemont, IL, 60439, USA}\\
\IEEEauthorblockA{\IEEEauthorrefmark{9}Department of Physics, Boston University, 590 Commonwealth Avenue, Boston, MA, 02215, USA}\\
\IEEEauthorblockA{\IEEEauthorrefmark{10}Harvard-Smithsonian Center for Astrophysics, 60 Garden Street, Cambridge, MA, 02138, USA}\\
\IEEEauthorblockA{\IEEEauthorrefmark{11}Steward Observatory and Department of Astronomy, University of Arizona, 933 N. Cherry Ave., Tucson, AZ 85721, USA}\\
\IEEEauthorblockA{\IEEEauthorrefmark{12}Three-Speed Logic, Inc., Victoria, B.C., V8S 3Z5, Canada}\\
\IEEEauthorblockA{\IEEEauthorrefmark{13}SLAC National Accelerator Laboratory, 2575 Sand Hill Road, Menlo Park, CA, 94025, USA}\\
}



\maketitle

\begin{abstract}
The South Pole Telescope Shirokoff Line Intensity Mapper (SPT-SLIM) is a millimeter-wavelength line-intensity mapping experiment, which was deployed on the South Pole Telescope (SPT) during the 2024-2025 Austral summer season. This pathfinder experiment serves to demonstrate the on-sky operation of multi-pixel on-chip spectrometer technology. We report on the cryogenic performance of the SPT-SLIM receiver for the first year of commissioning observations. The SPT-SLIM receiver utilizes an Adiabatic Demagnetization Refrigerator (ADR) for cooling the focal plane of superconducting filterbank spectrometers to a temperature of 150~mK. We demonstrate stable thermal performance of the focal plane module during observations consistent with thermal modeling, enabling a cryogenic operating efficiency above 80\%. We also report on the receiver control system design utilizing the Observatory Control System (OCS) platform for automated cryogenic operation on the SPT. 
\end{abstract}


\FloatBarrier

\section{Introduction}
\IEEEPARstart{L}{ine} intensity mapping (LIM) is an emerging technique for surveying large-scale structure by observing the redshifted emission of atomic or molecular lines both spatially and spectrally \cite{2018PhRvD..98d3529K}. The cosmological information available from a LIM survey is similar to a galaxy survey, but the technique can in principle scale to higher redshifts, and measurements of multiple lines can be cross-correlated to mitigate observational systematics \cite{2019BAAS...51c.101K}.

SPT-SLIM is a pathfinder experiment for demonstrating the use of on-chip millimeter-wavelength (mm-wave) spectrometers for LIM, and was deployed on the 10-meter South Pole Telescope (SPT) during the 2024-2025 Austral summer season. The SPT is a submillimeter-quality telescope with arcminute-scale resolution, primarily performing surveys of the cosmic microwave background (CMB) \cite{2011PASP..123..568C}. The South Pole site offers unparalleled sensitivity to the mm-wave sky for ground-based observations, with exceptionally high atmospheric transmission due to its altitude (2835~m) and low precipitable water vapor \cite{2011RMxAC..41...87R}.\IEEEpubidadjcol

The SPT-SLIM mm-wave on-chip filter bank spectrometers use a feed horn-coupled orthomode transducer (OMT) and microstrip to feed a bank of narrow-band filters. Each filter is instrumented with a lumped-element kinetic inductance detector (leKID), resulting in a compact, planar form factor with high KID density \cite{2022JLTP..209..879B, 2023ITAS...3359919C}. The SPT-SLIM camera contains a 9-pixel focal plane for observing redshifted CO (0.5~\lt~z~\lt~2) in the 150~GHz atmospheric window between the oxygen and water lines \cite{2022JLTP..209..758K}. The detectors are read out using the RF-ICE system \cite{2022SPIE12190E..24R}. Calibration and characterization of the detectors is discussed further in the associated proceedings \cite{Fichman_LTD25, Dibert_LTD25}. The Al KIDs used in SPT-SLIM require an operational temperature of 150~mK to maximize sensitivity. At the same time, SPT-SLIM needs to fit inside a constrained space within the SPT receiver cabin. In these proceedings, we present the cryostat receiver design and cryogenic performance.

\section{Receiver Design}
\begin{figure*}[t!]
\centering
\includegraphics[width=\textwidth]{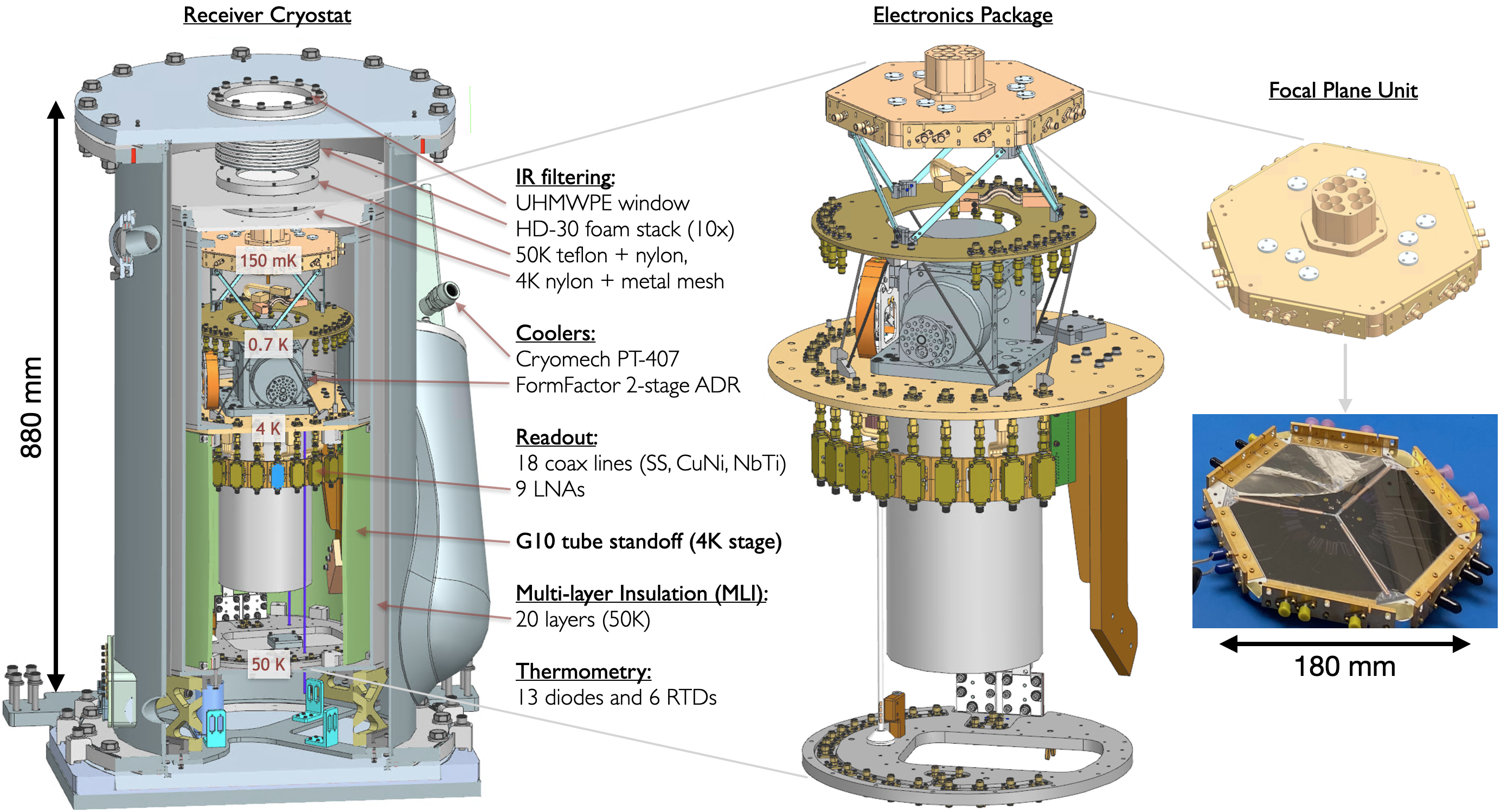}
\caption{Cross-section of the SPT-SLIM receiver cryostat design, electronics package insert, and focal plane unit. Cooling the focal plane to 150~mK is achieved using an ADR, backed by a PT407 cryocooler. The electronics package can be assembled independently before installing within the cryostat.}
\label{fig_cryostat_schematic}
\end{figure*}

The SPT-SLIM receiver utilizes a compact cryostat design, enabling it to be installed in the SPT receiver cabin alongside the SPT-3G camera. The cryostat was designed, built, and commissioned at Fermilab prior to deployment to the South Pole. A cross-section of the cryostat design is shown in Figure~\ref{fig_cryostat_schematic}.

\textbf{Cryogenics}: Sub-Kelvin temperatures are achieved via a FormFactor 155 2-stage adiabatic demagnetization refrigerator (ADR), with a Cryomech PT407 pulse tube cryocooler (PTC) used for cooling internal stages to 50~K and 4~K. The baseline cooling capacity of the ADR is 120~mJ at the FAA (150~mK) stage and 1.2~J at the GGG (0.7~K) stage, while the PT407 has a cooling capacity of 25~W at 55~K and 0.7~W at 4.2~K, sufficient for the compact size of the SPT-SLIM receiver. The PTC is mounted to the side of the cryostat and offset by 40\degree\ relative to the receiver optics, such that the PTC orientation remains close to vertical (with respect to gravity) in order to maximize cooling power at typical telescope observing elevations \cite{2021Cryo..11703323T}. The internal ADR is mounted to the 4~K stage of the cryostat, featuring a mechanical heat switch for coupling the sub-Kelvin stages to the 4~K bath during magnetization cycles. ADR hold time and the resulting duty cycle are largely driven by the thermal load on each sub-Kelvin stage, as well as the 4~K stage bath temperature during magnetization cycles. As such, several aspects of the cryostat mechanical design were aimed at reducing thermal gradients to the ADR (discussed further in Section~\ref{sec:performance}).

\begin{figure*}[t!]
\centering
\includegraphics[trim={0 10.5cm 0 0},clip=true,width=\textwidth]{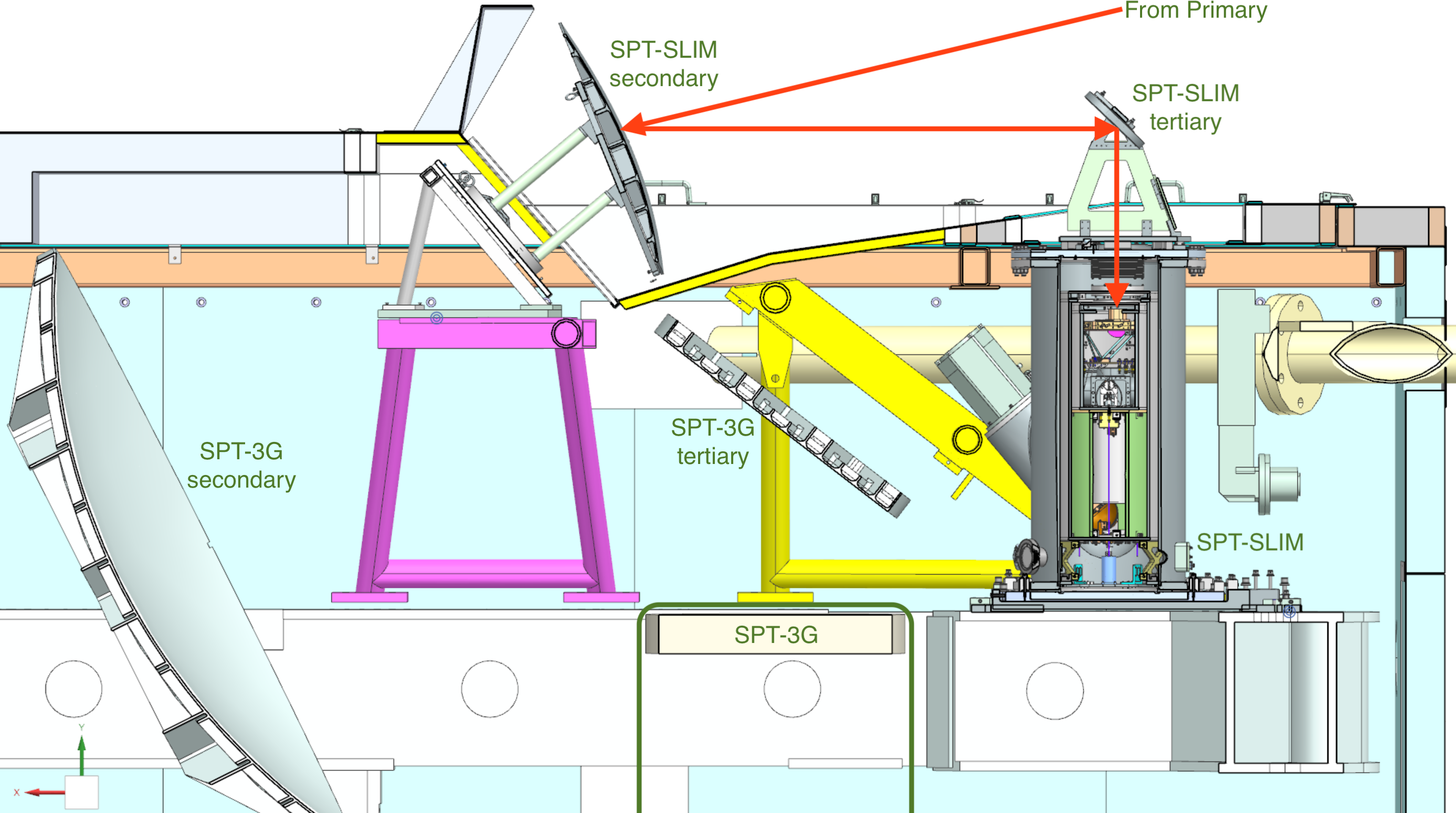}
\caption{Partial cross-section of the SPT receiver cabin, showing the SPT-SLIM cryostat mounting location and auxiliary mirrors for diverting the main SPT beam into the receiver.}
\label{fig_cabin}
\end{figure*}

\textbf{Optical design}: The SPT receiver cabin primarily houses the flagship SPT-3G instrument \cite{2022ApJS..258...42S}, with a small available area that has previously been used by the Event Horizon Telescope (EHT) South Pole receiver \cite{2018SPIE10708E..2SK}. The optical design for SPT-SLIM adopts a similar approach to EHT, using two additional custom mirrors that divert the main SPT beam and redirect it into the SPT-SLIM receiver located in the same space (shown in Figure~\ref{fig_cabin}). These secondary optics, along with the SPT-SLIM receiver, can be easily installed from the topside of the receiver cabin while the telescope is docked. This enables an observing configuration with minimal disruption to SPT-3G, although the experiments cannot run concurrently since the main beam is diverted.
Light enters the cryostat from the top flange, passing through an ultra-high-molecular-weight polyethylene (UHMWPE) vacuum window and several internal IR filters before terminating at the focal plane array. The IR filter stack consists of a radiatively cooled 10-layer HD-30 foam stack, a polytetrafluoroethylene (PTFE) and nylon filter at 50~K, and a nylon and metal-mesh low-pass filter at 4~K.

\textbf{Mechanical design}:
Each stage within the cryostat is mechanically supported from below, moving from the 300~K outer shell to the 150~mK focal plane module. The cross-sectional area of each support is optimized to minimize thermal loading to cooler stages while also minimizing vibrational modes. The 50~K stage is supported by four G10 trusses. The 4~K stage is supported by a 0.125~inch thick G10 tube that encapsulates the volume below the 4~K stage, also effectively reducing the radiative load on the 4~K stage and ADR housing. The GGG stage is supported by six 2.5~mm OD carbon fiber legs, and the FAA stage is supported by six 1~mm thick titanium (Ti~15-3-3-3) trusses. The predicted thermal loading on each stage is presented in Table~\ref{tab:thermal_performance}. The 50~K, 4~K, and sub-Kelvin stage assembly (and associated cabling) is referred to as the `electronics package,' and is designed such that the package can be assembled and warm-tested on a benchtop prior to installation within the cryostat.

\newlength{\sameheight}
\setlength{\sameheight}{0.73\columnwidth} 
\begin{figure}[!t]
\centering
\subfloat{\includegraphics[height=\sameheight]{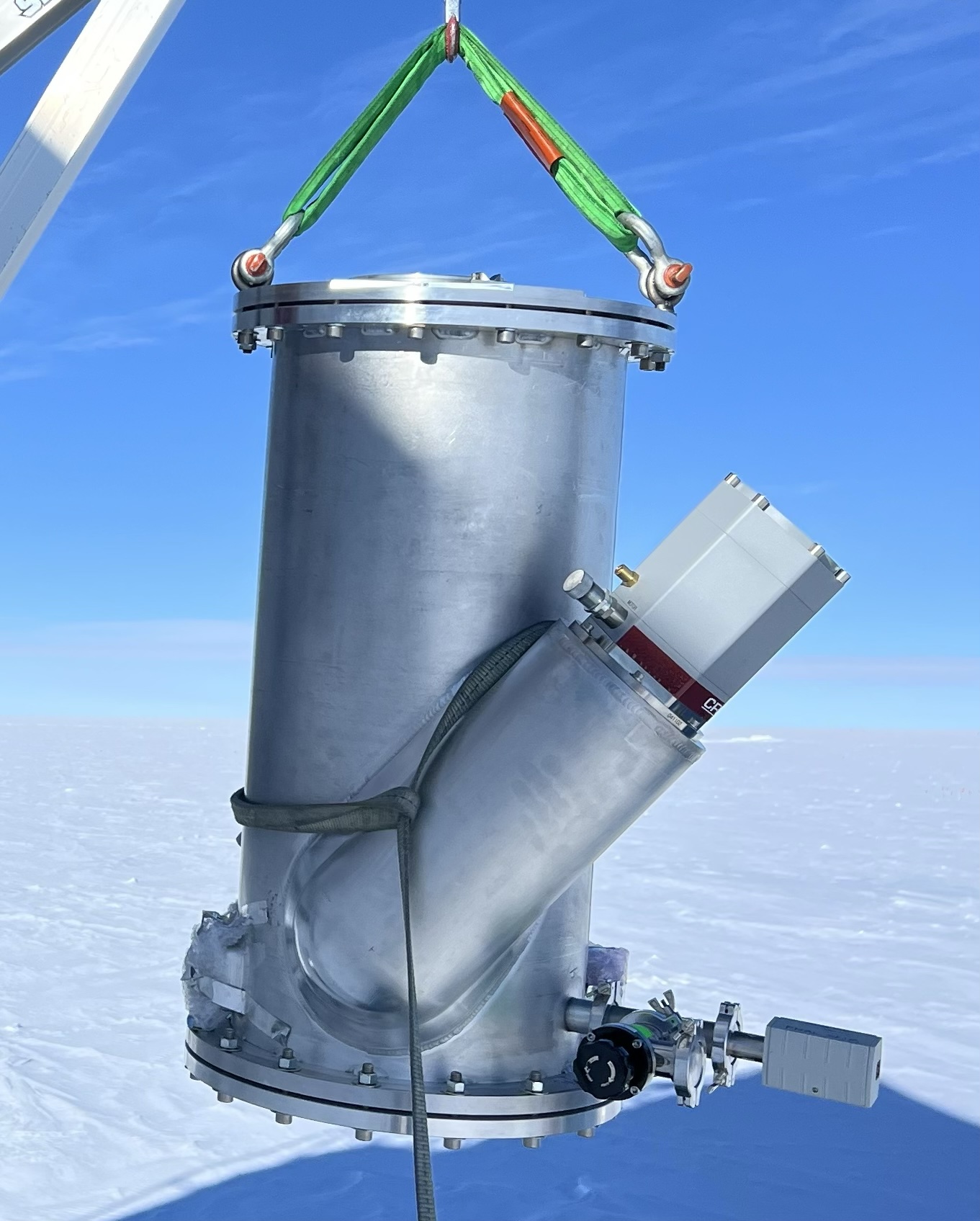}%
\label{fig_cryostat_photo}}
\hfill
\subfloat{\includegraphics[height=\sameheight]{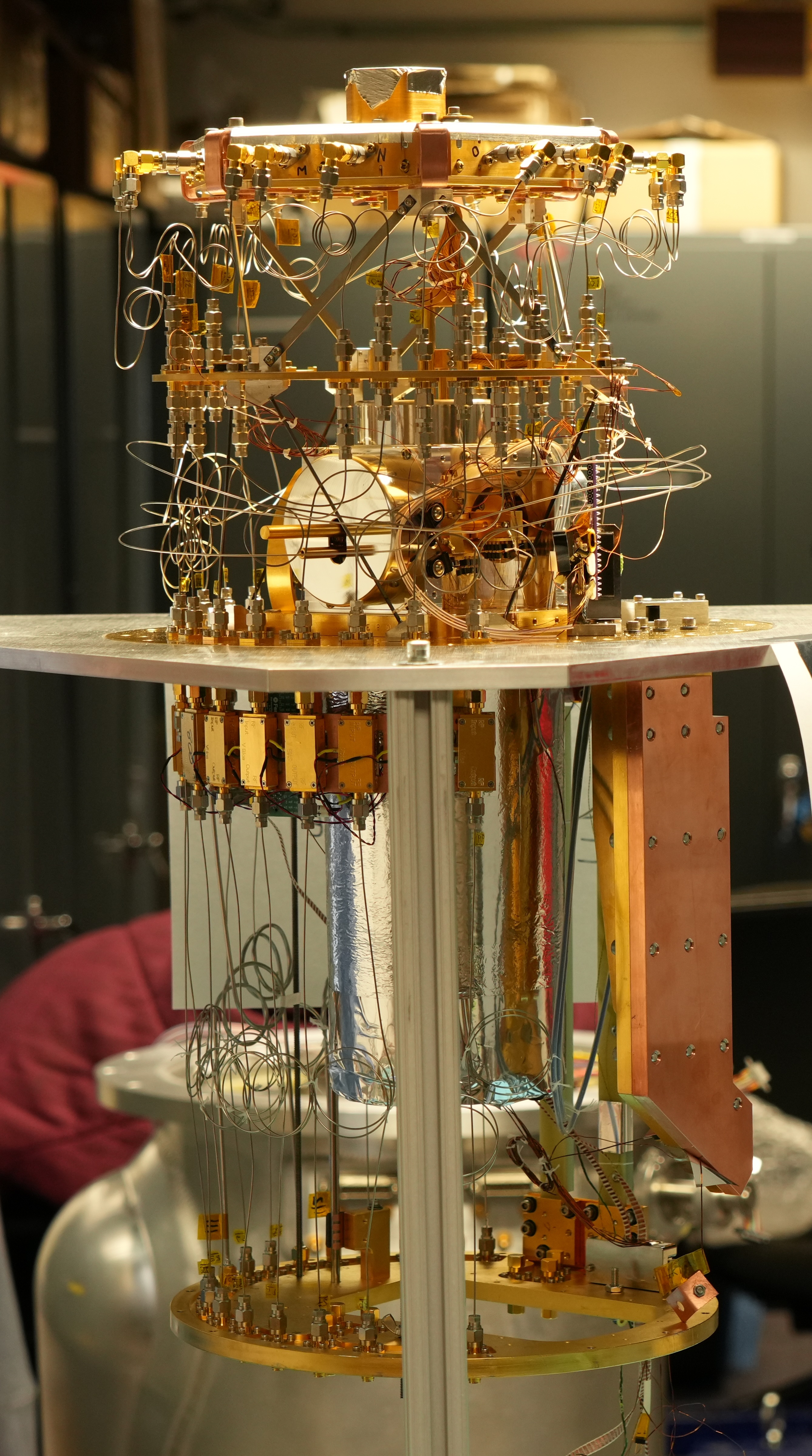}%
\label{fig_electronics_insert}}
\caption{\textit{Left}: The SPT-SLIM cryostat during installation on the SPT. The PTC angle is offset by 40\degree\ to account for typical telescope elevation tilt during observations. \textit{Right}: The electronics package insert, containing the ADR, LNAs, focal plane package, and associated readout lines.}
\label{fig_cryostat_photos}
\end{figure}

\textbf{Cryogenic Readout}:
The SPT-SLIM receiver houses 18 readout lines in total, to accommodate one input and output line for each of the 9 pixels on the focal plane array. The baseline design uses 0.86~mm OD cryogenic coaxial (`coax') cables manufactured by CryoCoax, with SMA connectors for interconnecting between cryostat stages. Stainless steel (SS) coax is used for all input lines up to the focal plane module, with Niobium Titanium (NbTi) coax used for the output lines through to 4~K, Cupronickel (CuNi) coax from 4~K to 50~K, and finally SS coax from 50~K to 300~K. In-lab testing prior to deployment revealed that a significant fraction of the SS coax lines failed at cryogenic temperatures due to thermal contraction within the crimped SMA connectors. For quality control, we performed dunk tests in liquid nitrogen for each SS coax line prior to installation. During deployment, four SS coax input lines between 50~K and 4~K were replaced with 2.19~mm OD lines with soldered SMA connections fabricated by Fermilab.

Each of the 9 readout lines features a cryogenic low-noise amplifier (LNA) coupled beneath the 4~K stage, produced by Arizona State University with a flat 30~dB gain from 0.5 to 3~GHz. A custom low-noise power supply using Analog Devices LT3042 linear regulators was designed to bias each LNA, drawing 7~mA at 1.5~V during operation at 4~K. One LNA was removed during deployment as a result of ESD damage. The additional 0.1~W of loading deposited by the LNAs on the 4~K stage during operation is not accounted for in Table~\ref{tab:thermal_performance}, as the LNAs remained unpowered during ADR magnetization cycles (discussed further in Section~\ref{sec:control}).

The cryostat also includes extensive thermometry, with 13 Lakeshore DT-670 silicon diodes used for monitoring various points along the 50~K and 4~K assemblies, and 6 Lakeshore ruthenium oxide (Rox) thermometers for sub-Kelvin monitoring and ADR temperature regulation.

\setlength{\sameheight}{0.82\columnwidth} 
\begin{figure*}[!t]
\centering
\subfloat{\includegraphics[height=\sameheight]{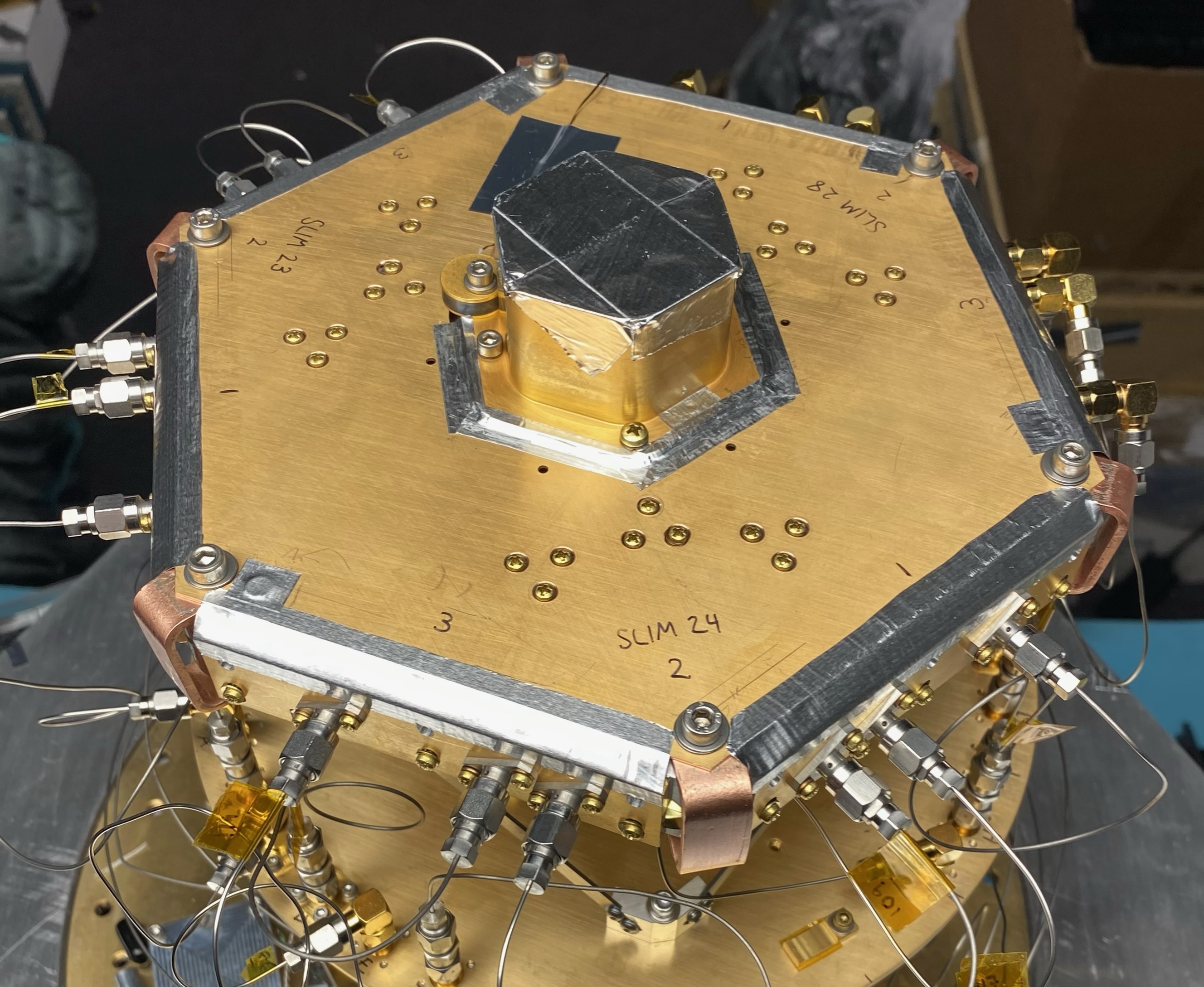}%
\label{fig_focalplane_top}}
\hfill
\subfloat{\includegraphics[height=\sameheight]{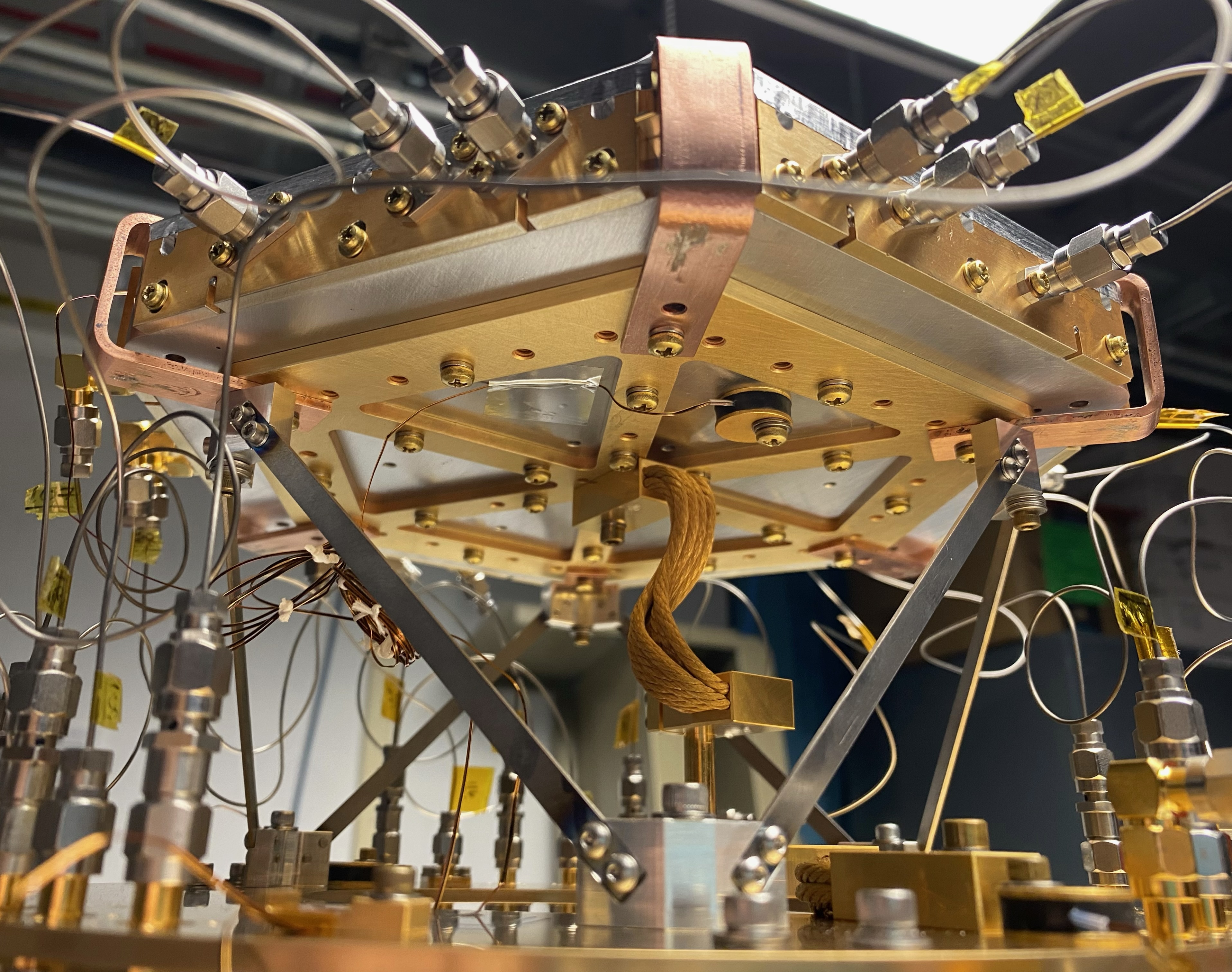}%
\label{fig_focalplane_bottom}}
\caption{\textit{Left}: Topside of the focal plane package, featuring a gold-plated copper shield above the aluminum detector module box. The shield is mounted via copper tabs to the underside. \textit{Right}: Underside of the focal plane package, showing the gold-plated copper spiderweb mount for the detector module box. These features were highly effective in reducing loading and thermal gradients across the detector module while also minimizing additional thermal mass.}
\label{fig_focalplane}
\end{figure*}

\begin{table*}[!t]
\renewcommand{\arraystretch}{1.3}
\caption{SPT-SLIM cryostat thermal loading summary.}
\label{tab:thermal_performance}
\centering
\small 
\begin{NiceTabular}{|
  >{\centering\arraybackslash}m{1.2cm}|
  >{\centering\arraybackslash}m{1.7cm}
  >{\centering\arraybackslash}m{1.7cm}
  >{\centering\arraybackslash}m{1.7cm}
  >{\centering\arraybackslash}m{1.7cm}
  >{\centering\arraybackslash}m{2.1cm}
  !{\vrule width 0.3pt}
  >{\columncolor{myblue}\centering\arraybackslash}m{2.1cm}
|}%
\Hline
\Block{2-1}{\textbf{Stage}} &
\multicolumn{5}{c!{\vrule width 0.3pt}}{\textbf{Modeled load}} &
\cellcolor{mywhite}\textbf{Measured load} \\
\cline{2-6}\cline{7-7}
& \textbf{Conductive}
& \textbf{Wiring}
& \textbf{Radiative}
& \textbf{Optical}
& \cellcolor{headgray}\textbf{Total}
& \cellcolor{myblue}\textbf{Total} \\
\Hline
50K & 2.65~W & 0.13~W & 11.5~W & 0.25~W & \cellcolor{mygray}14.5~W & 15~W \\
4K  & 101~mW & 136~mW & 7~mW   & 321~mW  & \cellcolor{mygray}113~mW  & 110~mW \\
GGG & 9.33~$\mu$W & 5.30~$\mu$W & 0.21~$\mu$W & 0~$\mu$W & \cellcolor{mygray}14.8~$\mu$W & 17.0~$\mu$W \\
FAA & 0.93~$\mu$W & 0.28~$\mu$W & 0.42~$\mu$W & 0.01~$\mu$W & \cellcolor{mygray}1.62~$\mu$W & 2.20~$\mu$W \\
\Hline
\end{NiceTabular}
\end{table*}

\section{Cryogenic Performance}
\label{sec:performance}

The cryogenic performance of the SPT-SLIM receiver throughout deployment was largely in agreement with model expectations, as shown in Table~\ref{tab:thermal_performance}. The total thermal loading is broken down into several subcategories: conductive loading through mechanical supports, conductive loading through wiring for the detector readout along with ADR magnet and LNA bias lines, radiative loading from surfaces within the cryostat, and optical loading from light passing through the window and absorbed by the IR filtering. The measured load on the 50~K and 4~K stages is derived from the PT407 capacity curve and base temperatures of each PTC head, measuring 41~K and 2.7~K respectively. The GGG-stage temperature increases over the course of an ADR regulation cycle, where the loading on this stage can be derived from the GGG cooling capacity (1.2~J from base temperature to 1~K) and the measured temperature gradient. Finally, we infer the total FAA-stage load by dividing the available cooling capacity at the start of the 150~mK hold, corrected for the starting field, the fraction of the capacity used to cool the stage, and the ADR cycle parameters (including the magnet-soak bath temperature), by the observed regulation time. This measured load is slightly higher than the modeled load, likely due to uncertainties in the sub-Kelvin thermal conductivity of the coax and the titanium trusses, where literature-derived values were used in place of direct thermal conductivity measurements for these components \cite{2014JLTP..176..201K, 2019EPJQT...6....2K}.

The hold time of the ADR and the 4~K and 50~K stage temperatures were improved between in-lab testing and deployment due to several modifications that were made to the design. The multi-layer insulation (MLI) on the 50~K stage was doubled from 10 to 20 layers, along with 10 layers added to the G10 tube and other exposed surfaces between the 50~K and 4~K stages. The volume of heat straps and bus bar connecting the PTC heads to their connected stages was also doubled, reducing the previously measured thermal gradients at each interface. Optimizing the PTC frequency to 1.51~Hz also reduced the 4~K stage temperature by 0.1~K. The resulting 4~K stage temperature (and ADR magnet bath temperature) was 3.2~K, an improvement of 0.3~K compared to pre-deployment measurements.

The focal plane module assembly housing the detector arrays is fabricated from aluminum 6061 due to the ease of machining high-tolerance features such as feedhorns and waveguides, although aluminum alloys exhibit an exceptionally low thermal conductivity at 150~mK. In-lab testing showed moderate thermal gradients across the assembly and sub-par regulation hold times. To mitigate these issues, a gold-plated copper shield was added above the focal plane, along with a gold-plated copper spiderweb mount for coupling to the FAA head, both shown in Figure~\ref{fig_focalplane}. The gold-plated shield acts to reduce the radiative load, and is directly mounted to the spiderweb mount via copper tabs in order to couple this load directly to the FAA head rather than through the aluminum module. This shielding method served as an alternative to gold-plating the entire focal plane module, maintaining a separation between the lossy gold-plating and KIDs inside the assembly. The copper spiderweb mount acts to improve the heat-sinking of the module to the FAA head via increased surface area while minimizing thermal mass. We also note that the FormFactor ADR system houses internal thermometry with spooled cabling secured by GE varnish, which produced a thermal short when the varnish failed, causing the cable to unspool. Additional varnish was added during deployment to mitigate this failure mode. 

\begin{figure}[t!]
\centering
\includegraphics[width=\columnwidth]{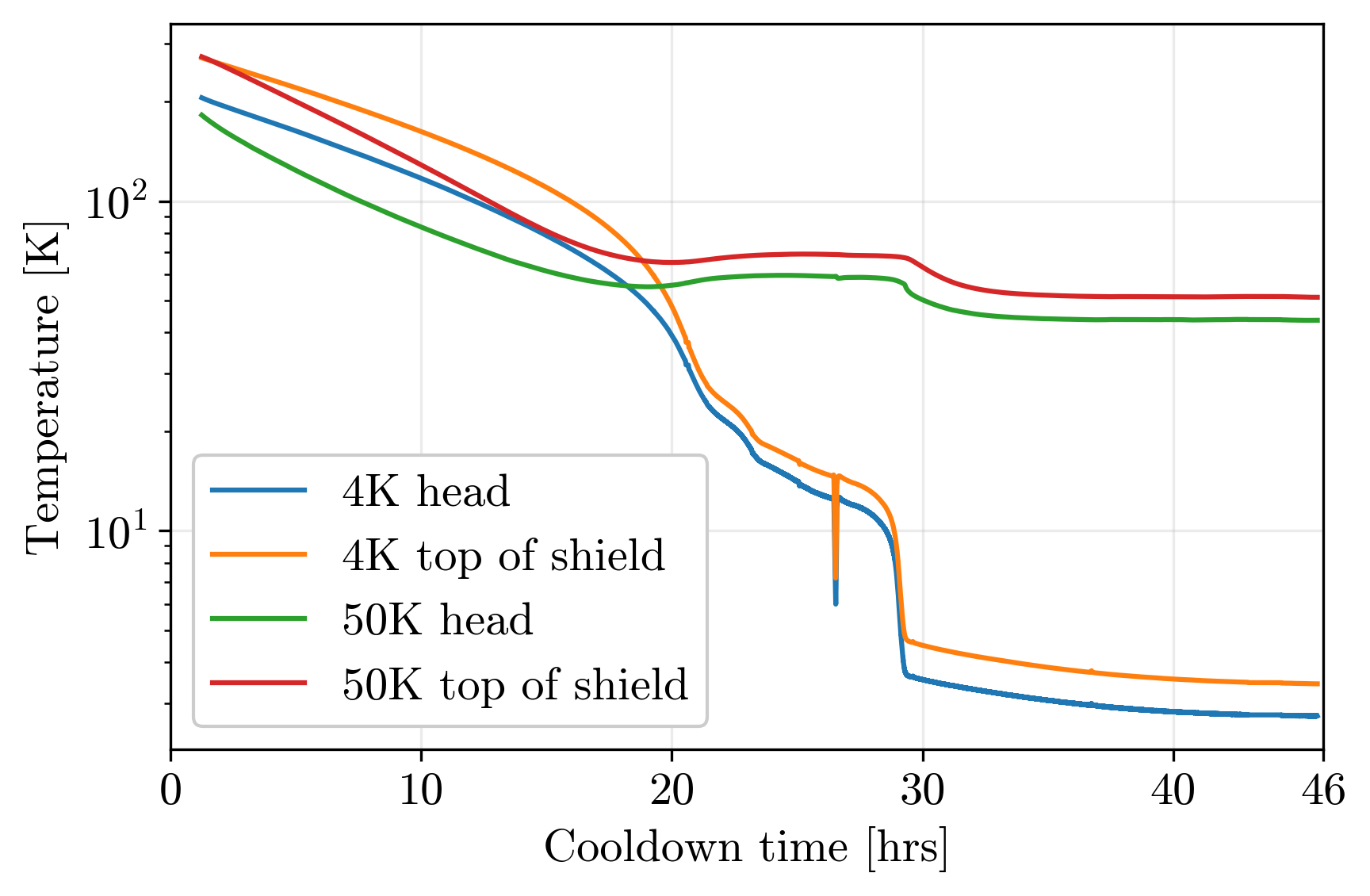}
\caption{Deployment cooldown curve of the SPT-SLIM cryostat, taking 46 hours to reach base temperature for ADR operation. The sharp feature at 26~hrs is a result of exercising the heat switch, briefly decoupling the \mbox{sub-Kelvin} stages from the PTC.}
\label{fig_cooldown}
\end{figure}

The deployed SPT-SLIM receiver achieved a hold time of 7 hours at 150~mK when running a 95-minute ADR magnetization cycle (20~minute ramp up, 30~minute soak, 45~minute ramp down), corresponding to an operational efficiency of 81\%. A thermometer on the underside of the module box (shown in Figure~\ref{fig_focalplane}) monitors the detector wafer bath temperature, measuring 160 mK with a stability of \lt 0.1~mK while observing. The module feed horn measured 155~mK during operation, demonstrating the focal plane package is isothermal to within the Rox thermometer accuracy of ±15~mK. The initial cooldown of the deployed SPT-SLIM receiver took 46~hours to reach base temperature for ADR operation, as shown in Figure~\ref{fig_cooldown}.

\section{Receiver Control System}
\label{sec:control}
\begin{figure}[t!]
\centering
\includegraphics[width=\columnwidth]{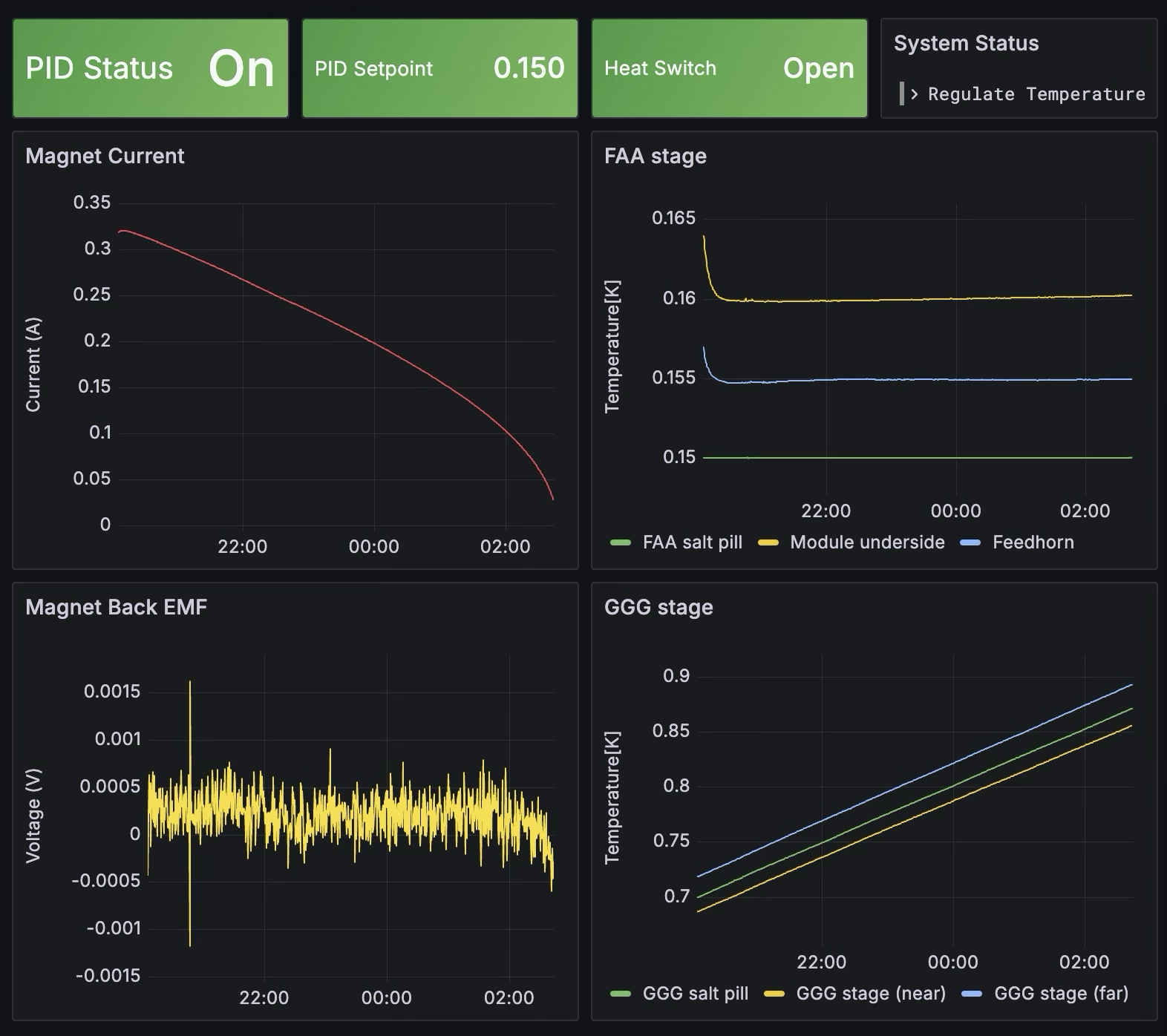}
\caption{Screenshot of ADR housekeeping data taken during 150~mK temperature regulation. OCS and Grafana were used for system monitoring and control.}
\label{fig_ocs}
\end{figure}

The receiver monitoring and control system was implemented using the Observatory Control System (OCS) platform\footnote{\url{https://github.com/simonsobs/ocs}}. A custom OCS plugin was created for interacting with the SPT-SLIM hardware, with Grafana and InfluxDB providing real-time monitoring of housekeeping data (demonstrated in Figure~\ref{fig_ocs}). The control system enabled automated cycling and regulation of the ADR with full control over system parameters, and included a switched PDU for powering the LNAs. We found that powering the LNAs immediately after the heat switch has opened (at which point the FAA and GGG stages are decoupled from the 4~K stage), produced a negligible impact on the regulation hold time despite heating the 4~K stage due to electrothermal power dissipation. Additional scripts were used to interface between the SPT-SLIM OCS agents and the SPT Generic Control Program (GCP) in order to automate cryogenic cycling and telescope observing schedules. 

\section{Conclusion}

SPT-SLIM has demonstrated the use of mm-wave on-chip spectrometers having successfully commissioned the receiver cryostat, operated it on the SPT, and performed observations of bright astrophysical sources. The cryogenic performance is in agreement with model expectations, enabling an observing efficiency of up to 81\% at 150~mK with $<$ 0.1~mK temperature stability during routine observing. The OCS platform was used for cryogenic monitoring and control, enabling automated observing with the SPT.

\section*{Acknowledgments}
SPT-SLIM is supported by the National Science Foundation under Award No. AST-2108763. The South Pole Telescope program is supported by the National Science Foundation (NSF) through awards OPP-1852617 and OPP-2332483. Work at Argonne, including use of the Center for Nanoscale Materials, an Office of Science user facility, was supported by the US Department of Energy, Office of Science, Office of Basic Energy Sciences and Office of High Energy Physics, under Contract No. DE-AC02-06CH11357. This document was prepared by the SPT-SLIM collaboration using the resources of the Fermi National Accelerator Laboratory (Fermilab), a U.S. Department of Energy, Office of Science, Office of High Energy Physics HEP User Facility. Fermilab is managed by FermiForward Discovery Group, LLC, acting under Contract No. 89243024CSC000002. The McGill team acknowledges funding from the Natural Sciences and Engineering Research Council of Canada and the Canadian Institute for Advanced Research, and the Canada Research Chairs program. This work is supported by UKRI Future Leaders Fellowship MR/W006499/1. Partial support is also provided by the Kavli Institute of Cosmological Physics at the University of Chicago. Support for this work for JZ was provided by NASA through the NASA Hubble Fellowship grant HF2-51500 awarded by the Space Telescope Science Institute, which is operated by the Association of Universities for Research in Astronomy, Inc., for NASA, under contract NAS5-26555. KSK was partially supported by an NSF Astronomy and Astrophysics Postdoctoral Fellowship under award AST-2001802, and by SLAC under award LDRD-24-008. We thank Heitor Mourato, Glenn Thayer, and Jose Velho in the BU Scientific Instrument Facility for assistance with mirror fabrication, and John Kovac and Miranda Eiben for assistance with anti-reflection coating optics.


\end{document}